\documentclass[reprint, superscriptaddress,
amsmath,amssymb,
aps, prl, 
floatfix,]{revtex4-2}

\usepackage{graphicx}
\usepackage{dcolumn}
\usepackage{bm}
\usepackage{lipsum}
\usepackage{xcolor}
\usepackage{siunitx}

\newcommand{\orange}[1]{\textcolor{black}{#1}}
\newcommand{\cyan}[1]{\textcolor{black}{#1}}

\newcommand{\magenta}[1]{\textcolor{black}{#1}}

\begin{document}
	
	\preprint{APS/123-QED}
	
	\title{Exploiting structural nonlinearity of a reconfigurable multiple-scattering system}
	
	\author{Yaniv Eliezer}
	\affiliation{Department of Applied Physics, Yale University, New Haven, Connecticut 06520, USA}

	\author{\magenta{Ulrich R\"{u}hrmair}}
	\affiliation{Physics Department, LMU M\"{u}nchen, 80539 M\"{u}nchen, Germany}
	
	\author{\cyan{Nils Wisiol}}
	\affiliation{Security in Telecommunications, Department of Software Engineering and Theoretical Computer Science, Technische Universit{\"a}t Berlin, Berlin, Germany}

	\author{\orange{Stefan Bittner}}
	\affiliation{\orange{Chair in Photonics, CentraleSup\'elec, LMOPS, Metz 57070, France}}
	\affiliation{\orange{Universit\'e de Lorraine, CentraleSup\'elec, LMOPS, Metz 57070, France}}
	
	\author{Hui Cao}
	\email{hui.cao@yale.edu}
	\affiliation{Department of Applied Physics, Yale University, New Haven, Connecticut 06520, USA}
	
	\date{\today}
	
	\begin{abstract}
		Nonlinear optics is a rapidly growing field that has found a wide range of applications. A major limitation, however, is the demand of high power, especially for high-order nonlinearities. Here, by reconfiguring a multiple-scattering system, we introduce `structural nonlinearity' via a nonlinear mapping between the scattering potential and the output light. Experimentally we demonstrate high-order, tunable \magenta{nonlinearities} at low power. The multiply-scattered light features enhanced intensity fluctuations and long-range spatial correlations. The flexibility, robustness and energy efficiency of our \magenta{approach} provides a versatile platform for exploring structural nonlinearities for various applications.
	\end{abstract}
	
	\maketitle
	
	
	

    Structural disorder and light scattering have recently been widely explored for photonic applications \cite{cao2022harnessing}. In most cases, the mapping from input to output fields is linear. However, nonlinear mapping is desired for various applications such as physical unclonable functions (PUFs) \cite{pavanello2021recent}, implementation of optical neural networks, neuromorphic computing and reservoir computing \cite{sunada2019photonic,wetzstein2020inference,rafayelyan2020large, Sunada20,marcucci2020theory}. Nonlinear mappings can be realized using nonlinear optical materials, which provide a nonlinear relation between input and output fields \cite{segev2013anderson, frostig2017focusing}. \cyan{The optical nonlinearity requires high light intensity, which is hard to achieve in random media due to spatial and temporal spreading of light by scattering}. Only low-order nonlinear processes such as second harmonic generation have been achieved in random scattering media \cite{Kravtsov_1991, tiginyanu_2000, melnikov_2004, Trull_2007, Faez_2009, Yao12}. Furthermore, a quadratic electro-optical process (Kerr effect) in a disordered system can induce temporal instability \cite{Kivshar_1992, SkipetrovMaynard_2000, SpivakZyuzin_2000, Sebbah_2011}. 
    
    \orange{
    Here we introduce another type of nonlinearity by exploring the nonlinear relation between a random potential and output light of a multiple-scattering system. Without using nonlinear optical materials, light scattering remains linear: the output field $E_o$ depends linearly on the input field $E_i$. The nonlinearity originates from multiple scattering, which can be described by the Born series \cite{born1926quantenmechanik}:
    \begin{equation}
        E_{o} = {\bf T} E_i = \left({\bf V} + {\bf V} {\bf G_0} {\bf V} + {\bf V}\left[{\bf G_0} {\bf V}\right]^2 + {\bf V}\left[{\bf G_0} {\bf {\bf V}}\right]^3  ...\right) E_{i} \nonumber
    \end{equation}
    where ${\bf T}$ is a matrix that captures the linear mapping from $E_i$ to $E_o$, ${\bf V}$ represents the scattering potential and ${\bf G_0}$ is the free-space Green's matrix. The first term in the expansion of ${\bf T}$ denotes single scattering (light scattered once), the second term double scattering, etc. With multiple scattering, the relation between the scattering potential configuration ${\bf V}$ and the output field $E_o$ is nonlinear. We term such nonlinearity as structural nonlinearity. The degree of structural nonlinearity increases with the number of scattering events (number of terms in ${\bf T}$ expansion). If single scattering dominates, the mapping from ${\bf V}$ to $E_o$ is approximately linear.
    }

    Such scheme is efficient in providing high-order nonlinearity at low power, and also avoids instability. However, the nonlinear mapping requires reconfiguring the scattering potential. Previous studies show that the refractive-index change induced by photorefractive effect is small \cite{freedman2006wave, schwartz2007transport}, and the change by thermo-optical effect is slow \cite{fleming2019perturbation, leonetti2014experimental}. 
    Dynamic coupling between multiply scattered light and colloidal particles can only statistically control the motion of colloidal particles \cite{douglass_2012, bianchi_2016}. In the \orange{microwave regime}, disordered cavities with active reconfigurable boundaries are explored for spatio-temporal focusing and power enhancement \cite{kaina2014shaping, Dupre_2015}, motion detection \cite{del2018dynamic}, analog computing \cite{Hougne_2018} and communication \cite{Hougne_2019, gros2021uncorrelated}. 
    
    In this work, we create an optical scattering cavity with a reconfigurable boundary. By tuning the cavity parameters, we are able to tune the order of structural nonlinearity. Consequently, fluctuations of local and total output light intensity are enhanced, and long-range spatial correlations are established. These effects will facilitate adaptive shaping of the scattering potential for light focusing and manipulation of total transmission \cite{resisi2020wavefront}.  The tunable, stable, and efficient structural nonlinearity opens the door to a broad range of applications including strong optical PUFs and nonlinear optical neural networks.   
	
	
	
\begin{figure}[hbtp]
		\centering
		\includegraphics[clip, trim = 0.0cm 0.0cm 0cm 0cm, width=1\linewidth]{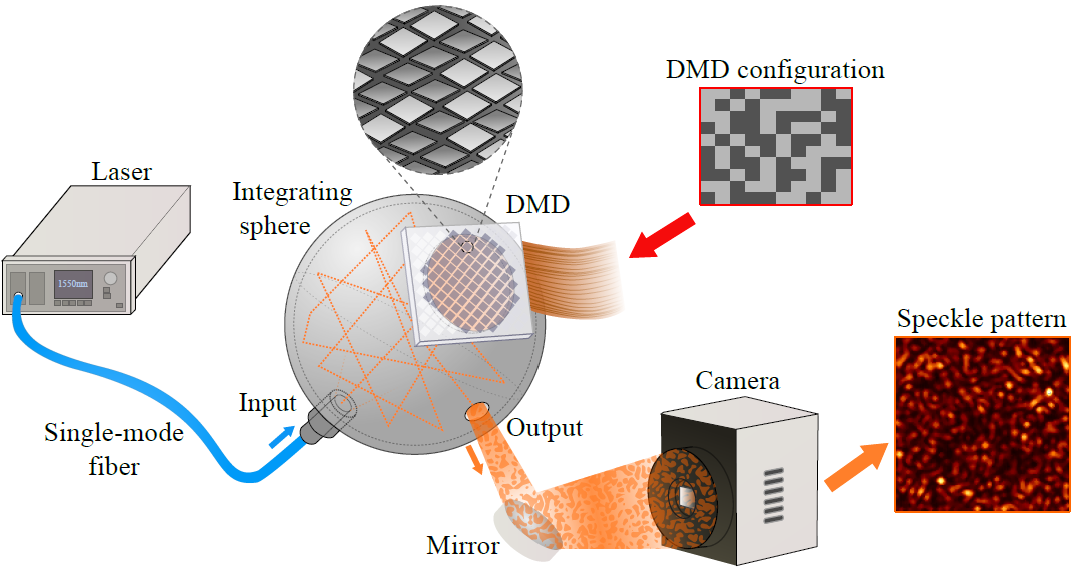}
		\caption{
			{Experimental setup}. A frequency-tunable continuous-wave fiber laser is coupled via a single mode fiber to an integrating sphere with an inner static rough boundary. A two-dimensional (2D) digital-mirror-array (DMD) covers one of the sphere's ports. Light is scattered multiple times inside the integrating sphere by its static boundary and the reconfigurable mirror array. Through a small open port, light leaks out of the cavity, and its intensity pattern is recorded by a digital camera. 
		}
		\label{fig:setup}
	\end{figure}

	As shown in Fig.~\ref{fig:setup}, our experimental setup is comprised of \cyan{an} integrating sphere (diameter \SI{3,75}{\centi\meter}). Its rough inner surface provides a static scattering potential. There are three ports on the sphere's boundary. The first port (diameter \SI{8}{\milli\meter}) is covered by a switchable digital mirrors device (DMD), which acts as a reconfigurable scattering potential. The DMD (Texas Instruments DLP9000X) consists of 2,560 $\times$ 1,600 micro-mirrors \cyan{(lateral dimension \SI{7,6}{\micro\meter}), each} can be flipped \cyan{to either} $+15^{\circ}$ or $-15^{\circ}$. A continuous-wave, linearly-polarized laser (Agilent 81940A) at wavelength $\lambda = $~\SI{1550}{\nano\meter} is coupled to a single mode fiber, which is connected to the second port of the integrating sphere. Input light through the second port is scattered multiple times inside the integrating sphere by the rough boundary and the switchable micro-mirrors. The third port (diameter \SI{3}{\milli\meter}) is left open, from which output light is directed by a mirror to an InGaAs camera (Xenics Xeva FPA-640). A linear polarizer is placed in front of the camera, which records the speckle intensity pattern for a specific DMD configuration. 
	
	At low input power (\SI{21}{\milli\watt}), the output field depends linearly on the input field, in spite of multiple scattering in the integrating sphere. However, the relation between the DMD configuration and the output speckle pattern is nonlinear, because light is scattered multiple times by the DMD. \magenta{In detail}, we divide the micro-mirror array into $M_p \times M_p$ macro-pixels. Each macro-pixel has only two states $+1$ and $-1$, in which all constituent micro-mirrors are tilted by $+15^{\circ}$ and $-15^{\circ}$, respectively. The DMD macro-pixels configuration can be described by a Boolean vector ${\textrm X}$ of length $M = M_p^2$ (total number of macro-pixels). There are $2^M$ possible configurations and each outputs a speckle pattern. The recorded two-dimensional intensity pattern is rearranged to an one-dimensional vector ${\textrm Y}$, \cyan{which is real (complex) for intensity (field)}. Its length $N$ is given by the number of camera pixels ($100 \times 100$) within the region of interest (including $\sim \, 20 \times 20$ speckle grains). The \cyan{exact} nonlinear mapping from the $k$-th DMD configuration ${\textrm X}^{(k)}$ to the corresponding output pattern ${\textrm Y}^{(k)}$ is obtained with a well-known technique of Boolean function analysis \cite{Kahn_1989, linial_1993, oDonnell_2014, pypuf}:
	\cyan{
	\begin{eqnarray}
	\label{eq:polynom}
    y_n^{(k)} &=& f\left(x_1^{(k)},...,x_{M}^{(k)}\right) \\
    &=& c^{(n,k)}_0 + \sum\limits_{m_1=1}^{M} c^{(n,k)}_{m_1} x_{m_1}^{(k)} + \sum\limits_{m_1=1}^{M} \sum\limits_{m_2=1}^{m_1} c^{(n,k)}_{m_1,m_2} x_{m_1}^{(k)} \, x_{m_2}^{(k)} \nonumber \\ 
    &+& \sum\limits_{m_1=1}^{M} \sum\limits_{m_2=1}^{m_1} \sum\limits_{m_3=1}^{m_2} c^{(n,k)}_{m_1,m_2,m_3}  x_{m_1}^{(k)} \, x_{m_2}^{(k)} \, x_{m_3}^{(k)} \quad ... \nonumber \\ 
    &+& c^{(n,k)}_{1,2,...,M} \, x_{1}^{(k)} \, x_{2}^{(k)} \, ... \, x_{M}^{(k)} \; , \nonumber
	\end{eqnarray}	
	}
	where $x_{m}^{(k)}$ is the $m$-th element of ${\textrm X}^{(k)}$, $y_n^{(k)}$ is the $n$-th element of ${\textrm Y}^{(k)}$, and $c^{(n,k)}_{\{\cdot\}}$ is the expansion coefficient, which is obtained by projection \cite{Supplementary}. 
	
	The number of $x_{m}^{(k)}$ factors in each term on the right-hand-side of Eq.~(\ref{eq:polynom}) gives the order $d$ of that term. $d = 1$ is the linear term, and $d \geq 2$ are nonlinear terms. Larger $d$ corresponds to high-order nonlinearity, and the maximal order is $d=M$. 
	We obtain the expansion coefficients in Eq.~(\ref{eq:polynom}) by fitting the measured output intensity patterns (subtracted by their average over $k$) for all possible DMD macro-pixel configurations \cite{Supplementary}. 
	Averaging $|c^{(n,k)}_{m_1, m_2, ... m_d}|$ for a certain order $d$, over $n$ and $k$, gives the mean expansion coefficient $\bar{c}(d)$ for this order. It is then normalized such that $\sum_d \bar{c}(d) = 1$. The effective order of nonlinearity is given by $\bar{d} \equiv \sum_d d \, \bar{c}(d)$.    

\begin{figure}[hbtp]
		\centering
		\includegraphics[clip, trim = 0.0cm 0.0cm 0cm 0cm, width=0.95\linewidth]{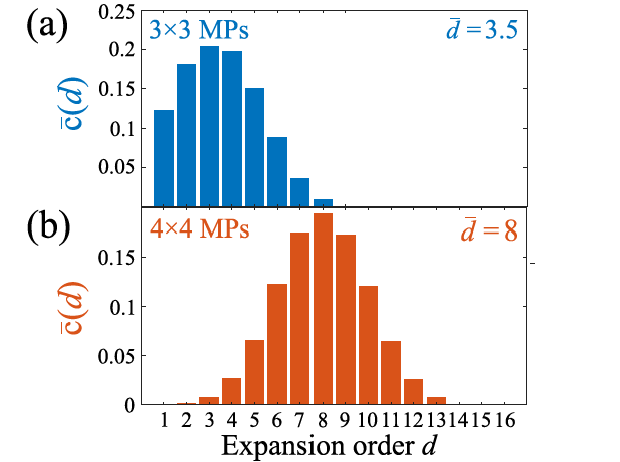}
		\caption{ {Experimentally measured structural nonlinearity}. 
			Expansion coefficient $\bar{c}$ vs.~order $d$ for output light intensity with $3\times3$ DMD macro-pixels in (a) and $4\times4$ in (b). \cyan{Note that the upper limit for $d$ is 9 in (a) and 16 in (b).} The distribution of $\bar{c}(d)$ shifts to higher order $d$ with larger number of macro-pixels. The effective order of nonlinearity $\bar{d}$ increases from $3.5$ (a) to $8.0$ (b).
		}
		\label{fig:exp}
	\end{figure}

	Figure~\ref{fig:exp}(a) shows the distribution of $\bar{c}(d)$ for $3\times3$ macro-pixels ($2^9$ configurations). $\bar{c}(d)$ spreads from $d = 1$ to $d=8$, reaching the maximum at $d = 3$. When the number of macropixels is increased to $4 \times 4$, the distribution of $\bar{c}(d)$ moves to larger $d$, with the maximum at $d = 8$. The mean $\bar{d}$ increases from $3.5$ to $8.0$, reflecting a higher order of nonlinearity.  As the total area of the DMD is fixed, the number of macro-pixels determines the degree of control of the scattering potential. Hence, the effective order of structural nonlinearity increases with the degree of control of the scattering potential. We further enhance the nonlinearity by using  $10 \times 10$ macro-pixels. There are $2^{100}$ possible configurations, which are impossible to measure and analyze to find the effective order of nonlinearity.
	
\begin{figure}[hbtp]
		\centering
		\includegraphics[clip, trim = 0.0cm 0.0cm 0cm 0cm, width=1\linewidth]{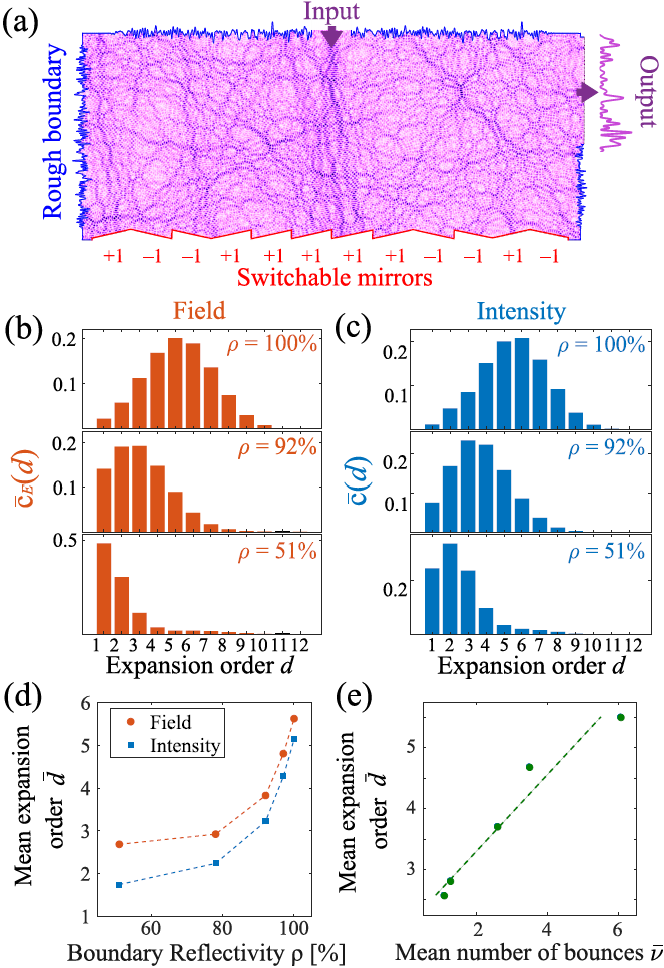}
		\caption{ {Tuning of structural nonlinearity}. 
			(a) Steady-state solution of light distribution inside a 2D cavity (dimension $130\lambda \times 60\lambda$) with static rough boundary (blue) and 12 switchable mirrors (red). Each mirror (length $10 \lambda$) is set independently to a tilt-angle of $+15^\circ$ (state of $+1$) or $-15^\circ$ (state of $-1$). Monochromatic light (wavelength $\lambda$) is injected from the input port (top), and exits through the output port (right), producing an 1D speckle pattern (containing $\sim70$ speckle grains). (b,c) Expansion coefficient $\bar{c}(d)$ vs. order $d$ for output field (b) and intensity (c). With increasing reflectivity $\rho$ of the cavity boundary, the distribution of $\bar{c}(d)$ moves to higher order $d$. 
			(d) Effective order of nonlinearity for output field $\bar{d}_E$ and intensity $\bar{d}$ grows with $\rho$. Since intensity is the square of field amplitude, it is slightly more nonlinear. 
			(e) Effective order of nonlinearity $\bar{d}$, scales with the mean number of bounces of optical rays off the mirror-array $\bar{\nu}$. Solid circles are numerical data for different boundary reflectivity $\rho$. Dashed line is a linear fit.  
		}
		\label{fig:nlordersim}
	\end{figure}

	We expect the strength of structural nonlinearity to depend on the number of times that light scatters off the DMD. Experimentally it is difficult to tune the number of scattering events, and we resort to numerical simulation. To save computation time, we consider a small two-dimensional (2D) cavity with rough boundary [Fig.~\ref{fig:nlordersim}(a)]. The number of micro-mirrors on the cavity boundary is $M=12$. Each mirror (with $100\%$ reflectivity) can be tilted by $+15^\circ$ (state of $+1$) or $-15^\circ$ (state of $-1$). Monochromatic light is injected through a small input port, and scatters off the rough boundary and micro-mirrors. To vary the number of scattering events at the switchable micro-mirrors, the optical reflectivity $\rho$ of the cavity boundary (away from the mirrors) is tuned between $51\%$ to $100\%$ to change the lifetime of light inside the cavity. Part of the light leaks out of an output port, and the spatial distributions of both field and intensity are computed in a full-wave simulation of light propagation inside the cavity for each possible configuration of the $M=12$ switchable mirrors ($2^{12}$ in total) \cite{Supplementary}. 
	Output field and intensity patterns are stored in two matrices ${\bf Y}_{\bf E}$ and ${\bf Y}_{\bf I}$ correspondingly. We map the mirror configurations matrix ${\bf X}$ to ${\bf Y}_{\bf E}$ and ${\bf Y}_{\bf I}$ using Eq.~(\ref{eq:polynom}), for different values of boundary reflectivity $\rho$. 
	
	Figure~\ref{fig:nlordersim}(b,c) shows the distribution of $\bar{c}_E(d)$ for output field, and that of $\bar{c}(d)$ for intensity. Both distributions shift to larger $d$ with increasing $\rho$. With low boundary reflectivity, the lifetime of light inside the cavity is short, and the number of bounces off the micro-mirror array is small. At $\rho = 51
	\%$, $\bar{c}_E(d)$ is peaked at $d=1$, thus structural nonlinearity is weak for output field [Fig.~\ref{fig:nlordersim}(b)]. Correspondingly, $\bar{c}(d)$ is peaked at $d=2$, indicating the nonlinearity for intensity is mostly from the square of field amplitude [Fig.~\ref{fig:nlordersim}(c)]. With increasing $\rho$, the effective order of nonlinearity for both field $\bar{d}_E$ and intensity $\bar{d}$ rises monotonously, despite $\bar{d}_E$ is slightly lower than $\bar{d}$ [Fig.~\ref{fig:nlordersim}(d)]. \orange{As the boundary reflectivity is higher, light stays longer inside the cavity and has more chance of scattering off the switchable mirrors. An increase in the number of scattering events leads to higher order structural nonlinearity}. 

\newpage

 	To quantify the number of bounces off the micro-mirror array, we conduct a classical ray tracing simulation in the same cavity with varying boundary reflectivity \cite{Supplementary}. 
 	Briefly, we launch many optical rays into the cavity, and trace individual ray propagation until it dies out. The number of bounces off the mirrors is weighted by the intensity of the ray at each bounce to give the effective number of bounces for a single ray. Then we average this number over all rays escaping through the output port, to obtain the mean number of bounces $\bar{\nu}$ for different $\rho$. \cyan{Fig.~\ref{fig:nlordersim}(e) shows that $\bar{d}$ scales linearly with $\bar{\nu}$, indicating that the average number of light bounces is closely related to the order of structural nonlinearity.}


	
\begin{figure}[hbtp]
		\centering
		\includegraphics[clip, trim = 0.0cm 0.0cm 0cm 0cm, width=1\linewidth]{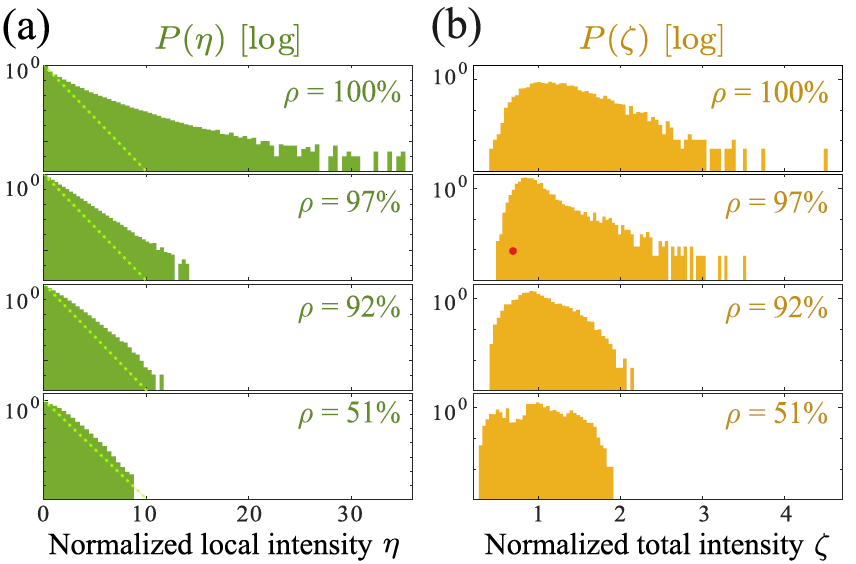}
		\caption{ {Output intensity fluctuations.} 
		(a) Probability density function (PDF) $P(\eta)$ of local intensity (normalized by the mean of all mirror configurations) $\eta$. With increasing boundary reflectivity $\rho$, $P(\eta)$ becomes broader and develops a tail. PDF $P(\tilde{\eta})$ of intensity normalized by the spatial-average for individual mirror configuration (dotted line) exhibits an exponential decay, regardless of $\rho$.
		(b) PDF of the total output intensity for individual mirror configuration, $P(\zeta)$, is widened and skewed with increasing $\rho$. Fluctuations of local and total output intensities are stronger with higher-order structural nonlinearities. All the results are obtained from numerical simulation of a 2D cavity with parameters identical to those in Fig.~\ref{fig:nlordersim}.
		}
		\label{fig:specklestatsim}
	\end{figure}

	Next we investigate how structural nonlinearity influences statistics of output light. Previously, `hot spots' of field intensities akin to rogue wave formation have been observed in random-scattering microwave and chaotic nano-photonic cavities \cite{hohmann2010freak, liu2015triggering}. Here, we explore the local intensity fluctuation with varying scattering potential. First, light intensity $I_k(r)$ \cyan{at a spatial position $r$} for a switchable-mirror configuration $k$ is normalized by its mean over all possible configurations $\langle I_k(r) \rangle_k$ as follows: $\eta_k(r) = I_k(r)/ \langle I_k(r) \rangle_k$. Then the probability density function for the normalized intensity $P(\eta)$ is obtained from the data at different output positions and mirror configurations. Figure~\ref{fig:specklestatsim}(a) shows $P(\eta)$ from the numerical simulation of 2D cavities with different boundary reflectivity $\rho$. As $\rho$ increases from $51\%$ to $100\%$, $P(\eta)$ develops a heavy tail at large $\eta$. The output intensity fluctuation grows with the number of scattering events at the switchable-mirror array. The intensity statistics gradually transforms from sub-Rayleigh to super-Rayleigh with increasing structural nonlinearity. 
	
	For comparison, we normalize the output intensity by the spatial-average $\langle I_k(r) \rangle_r$ for individual mirror configuration as follows: $\tilde{\eta}_k(r) = I_k(r)/ \langle I_k(r) \rangle_r$. As shown by the dotted line in Fig.~\ref{fig:specklestatsim}(a), the probability density function (PDF) $P(\tilde{\eta})$ exhibits an exponential decay for all values of $\rho$. Thus the speckle pattern for individual mirror realization satisfies the Rayleigh statistics \cite{Supplementary}. 
	
	The deviation of $P(\eta)$ from Rayleigh PDF results from the fluctuation of \orange{the total output intensity} from one mirror configuration to another. To illustrate this point, we compute the normalized \orange{total intensity} $\zeta_k = \int I_k(r) \, dr / \langle \int I_k(r) \, dr \rangle_r = \langle I_k(r) \rangle_r / \langle I_k(r) \rangle_{r,k}$, and plot its PDF in Fig.~\ref{fig:specklestatsim}(b). With increasing boundary reflectivity $\rho$, $P(\zeta)$ is broadened and skewed. \orange{The widening of $P(\zeta)$ indicates that the fluctuation of total output intensity with mirror configurations is enhanced by structural nonlinearity.}  
	
\begin{figure}[hbtp]
		\centering
		\includegraphics[clip, trim = 0.0cm 0.0cm 0cm -.5cm, width=1\linewidth]{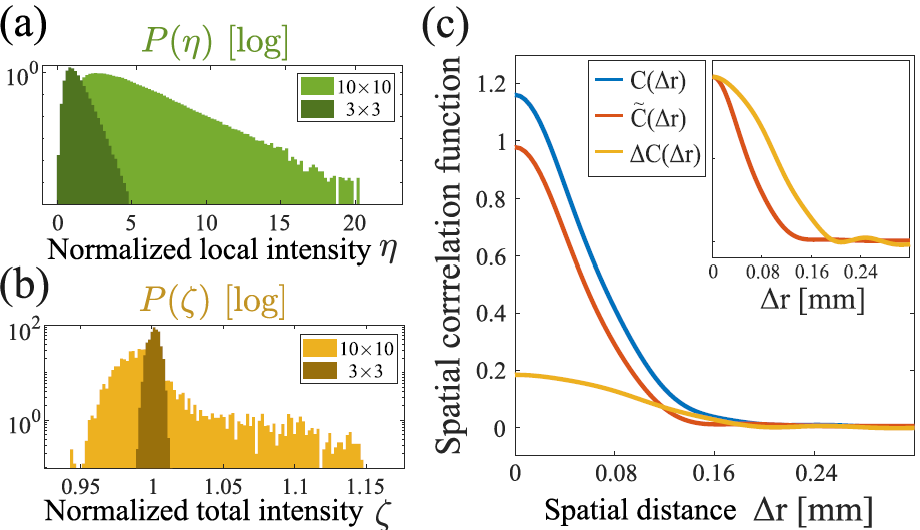}
\caption{ {Measured speckle statistics}.
		(a,b) Probability density function of local speckle intensity $\eta$ in (a) and of \orange{total} output intensity $\zeta$ in (b) is much broader with a heavy tail for $10 \times 10$ macro-pixels than that for $3 \times 3$ macro-pixels. 
		(c) Spatial correlation $C(\Delta r)$ of $\eta$ is stronger than that $\tilde{C}(\Delta r)$ of $\tilde{\eta}$. Their difference $\Delta C(\Delta r) = C(\Delta r) - \tilde{C}(\Delta r)$ (normalized by $\Delta C(0)$ in the inset) has a larger width than $\tilde{C}(\Delta r)$. Spatial correlation of local intensity fluctuation is enhanced with $10 \times 10$ macro-pixels.
		}
		\label{fig:C1C2}
	\end{figure}

	Experimentally we measure the intensity statistics of output speckle patterns from the integrating sphere. To increase the structural nonlinearity, we divide the DMD into $10\times10$ macro-pixels, and record the output speckle patterns for a large ensemble ($>10,000$) of random binary configurations of DMD macro-pixels.  Figure~\ref{fig:C1C2}(a) shows the local intensity PDF, $P(\eta)$, is much more extended than that for  $3\times3$ macro-pixels. Its heavy tail reflects stronger local intensity fluctuations for $10 \times 10$ macro-pixels. The extraordinary intensity values facilitate light focusing, i.e., enhancing local intensity of output light by optimizing the DMD configuration. Fig.~\ref{fig:C1C2}(b) shows the fluctuation of total output intensity $P(\zeta)$. For $3 \times 3$ macro-pixels, $P(\zeta)$ has a narrow, symmetric distribution around $\zeta =1$. It becomes much wider and skewed with a heavy tail for $10 \times 10$ macro-pixels. The broadening of $P(\zeta)$ increases the range of control of the total output intensity (transmittance) by manipulating the scattering potential with the DMD.     
	
	Finally, we show that structural nonlinearity also enhances spatial correlations of intensity fluctuations with scattering potentials. For that purpose, we evaluate the \orange{correlation of changes in output intensities at different locations with varying DMD configurations. With the local intensity normalized by its mean over different DMD configurations, the spatial correlation function is $C(\Delta r)~\equiv~\langle \eta_k(r) \, \eta_k(r + \Delta r) \rangle_{r,k} -1 $. For comparison, the intensity is normalized by the spatial average for a single DMD realization, and the corresponding correlation function is $\tilde{C}(\Delta r)~\equiv~\langle \tilde{\eta}_k(r) \, \tilde{\eta}_k(r + \Delta r) \rangle_{r,k} -1 $.} Figure~\ref{fig:C1C2}(c) shows that $C(\Delta r)$ exceeds $\tilde{C}(\Delta r)$ for $10 \times 10$ macro-pixels. Note that $C(0)$ is equal to the variance of $\eta$, which exceeds $\tilde{C}(0)$. The difference $\Delta C(\Delta r) \equiv C(\Delta r) - \tilde{C}(\Delta r)$ decays slower with $\Delta r$ than $\tilde{C}(\Delta r)$, revealing the increased range of correlation. In contrast, the difference between $C(\Delta r)$ and $\tilde{C}(\Delta r)$ is much smaller for $3 \times 3$ macro-pixels (not shown). This confirms that the enhanced correlation of intensity fluctuations results from high-order structural nonlinearity. 
	
    In conclusion, our method of achieving highly nonlinear response is robust, flexible and power efficient. Such tunable nonlinearity may be utilized, for example, to enhance the security of optical scattering PUFs \cite{pappu2002physical}, or to improve the performance of large-scale reservoir computing \cite{rafayelyan2020large}. We expect the current platform in particular and structural nonlinearities in general to have diverse applications in metrology, optical computing and cryptography.	
    
	
	\begin{acknowledgments}
		This work is supported by the US Air Force Office of Scientific Research (AFOSR) under Grant No.  FA9550-21-1-0039. The authors would like to thank 
		Seng Fatt Liew, 
		Brandon Redding, 
		Hasan Y{\i}lmaz, 
		KyungDuk Kim,
		SeungYun Han,    
		Liam Shaughnessy,
		Chen Chun-Wei,    
		Rasmus Rasche, 
		Michael Lachner, 
		Adomas Baliuka,
		Sara Nocentini, 
		Giuseppe Emanuele Lio
		and Francesco Riboli
		for valuable information and stimulating discussions.
		We acknowledge the computational resources provided by the Yale High Performance Computing Cluster (Yale HPC).
		\orange{S.\ B.\ acknowledges support for the Chair in Photonics from Minist\`ere de l'Enseignement Sup\'erieur, de la Recherche et de l'Innovation; R\'egion Grand-Est; D\'epartement Moselle; European Regional Development Fund (ERDF); Metz M\'etropole; GDI Simulation; CentraleSup\'elec; Fondation CentraleSup\'elec.}
	\end{acknowledgments}
	
	
	\bibliography{A_mainbib} 

\begin{thebibliography}{41}
\bibitem [{Sup()}]{Supplementary}%
  \BibitemOpen
  \bibinfo {note} {{S}ee {S}upplementary {M}aterial at [URL will be inserted by
  publisher] for further technical details, numerical simulations and other
  results}\BibitemShut {NoStop}%
\end{thebibliography}
	
\end{document}